\documentstyle[aas2pp4]{article}

\lefthead{Grimberg, Sadler, and Simkin}
\righthead{EELG-PKS0349-27}

\lefthead{Grimberg, Sadler, and Simkin}
\righthead{EELG-PKS0349-27}

\begin{document}

\title{Extended Emission Line Gas in Radio Galaxies - PKS0349-27} 

\author{B. I. Grimberg}
\affil{Michigan State University, East Lansing MI 48824-1116}
\authoremail{grimberg@pa.msu.edu}

\author{E. M. Sadler
\altaffilmark{1}} 
\affil{School of Physics, University of Sydney, NSW 2006, Australia}
\authoremail{EMS@Physics.usyd.edu.au}

\and

\author{S. M. Simkin
\altaffilmark{1}} 
\affil{Michigan State University, East Lansing, MI 48824-1116}
\authoremail{simkin@grus.pa.msu.edu}

\altaffiltext{1}{Visiting Astronomer, Cerro Tololo Inter-American
Observatory.  CTIO is operated by AURA, Inc.\ under contract to the
National Science Foundation.}

\slugcomment{submitted to Ap.J. January 03, 1999, Accepted March 15, 1999}

\begin{abstract} 
PKS0349-27 is a classical FRII radio galaxy with an AGN host which has
a spectacular, spiral-like structure in its extended emission line gas
(EELG\footnote{We have chosen this designation rather than the more
common abbreviation EELR(egions) because we wish to emphasize the
physical difference between this type of highly extended gas and the
extended narrow line regions around the nuclei of active galaxies.}).
We have measured the velocity field in this gas and find that it splits
into 2 cloud groups separated by radial velocities which at some points
approach 400 \,km\,s.$^{-1}$  Measurements of the diagnostic emission
line ratios [OIII]\,5007/H$\beta$, [SII]\, 6716+6731/H$\alpha$, and
[NII]\,6583/H$\alpha$ in these clouds show no evidence for the type of
HII region emission associated with starburst activity in either
velocity system. The measured emission line ratios are similar to those
found in the nuclei of narrow-line radio galaxies, but the extended
ionization/excitation cannot be produced by continuum emission from the
active nucleus alone.  We present arguments which suggest that the
velocity disturbances seen in the EELG are most likely the result of a
galaxy-galaxy collision or merger but cannot completely rule out the
possibility that the gas has been disrupted by the passage of a radio
jet.

\end{abstract}


\section{Introduction} 

PKS0349-27 is a prime example of a classical, double-lobed (FRII) radio
galaxy (\cite{Bo65,Ch77}) with a narrow emission line nuclear
spectrum.  It is also a spectacular example of a radio AGN host with an
extended gaseous component which has a structure reminiscent of a
barred spiral (\cite{Ha87,Ba88,Si89}; see Figure\,1b in
section~\ref{images}).  It has a redshift of $\sim$19800\,km\,s$^{-1}$
(\cite{Se68}), corresponding to a scale of
$\sim$0.96\,kpc\,$('')^{-1}h.^{-1}$ This galaxy appears to fit into the
``rotator'' class of FRII, double-lobed, radio galaxies defined by
\cite{Ba92}.

\subsection{Previous Observations} \label{prev}

Earlier kinematic studies of this object have suggested that its
centrally concentrated ionized gas is in rapid rotation as well as
expansion (with a rotation axis at $\sim$PA140-180$^{o}$,
\cite{Da84,Ta89}).  Its non-nuclear spectrum has been extensively
observed (op cit. and \cite{Ro87,SB96,Bi96})\, and incorporated in
several  analytical discussions of the type of ionization/excitation
conditions found in the non-nuclear regions of active galaxies (op
cit.  and \cite{Vi92,Wi97}).  On the basis of HST, UV continuum images,
\cite{Zi98} have concluded that there is no strong UV emission from its
nucleus; however, they do report some evidence for extended UV emission
in the $\sim$3'' region surrounding the nucleus.

Detailed radio continuum images of this object have been published by
\cite{Ek89} at 4.89 GHz and by \cite{Ba88} at 1400 MHz.

\subsection{EELG, Mergers, and Star Formation} \label{EELG}

Although it was initially thought to be a rare phenomenon, the
presence of EELG is quite common in many strong radio galaxy hosts (as
reviewed and discussed by \cite{Baa89}). It may well be related to
the ``alignment'' effect seen in more distant, powerful radio galaxies
(cf.  \cite{Mc95,Mc96,Be96}) and frequently seems to reflect a
different kinematic system from the stellar component of the radio
galaxy host (cf. the reviews in \cite{Baa89}, and \cite{Ta89}).

Ideas about the origins of this gas, its ionization/excitation
parameters, and kinematic properties are reviewed in the references
cited above  as well as \cite{Ko98}, and \cite{Ta98}. These can be
briefly summarized as follows:

\begin{enumerate}

\item Origins: The gas may be native, from stellar mass loss; acquired
in a ``cooling flow;'' accreted from primordial cold gas; or acquired
from a gas-rich galaxy in a merger or close encounter.

\item Ionization/excitation: The gas may be photoionized by a hard
continuum source originating in the nucleus; shock heated by
collisional impact due to radio jets or tidal disturbances from a
merger or close encounter and post shock, radiative re-heating;
photoionized by starburst activity; or a combination of these.

\item Kinematics:  The gas may be in stable orbits about the host;
disrupted by collisions arising from mergers and tidal disruption; or
disrupted by the impact of a radio jet.

\end{enumerate}

Nearby objects are particularly important for resolving the complex
interactions noted above because high spatial resolution is needed to
distinguish between the different physical zones which can arise within
the clouds.  The number of nearby, intrinsically powerful radio sources
is, however, limited  and thus it is important to subject those which
do exist to intense analysis.  This implies that these nearby objects
should be restudied as instrumental capabilities improve our ability to
measure such objects more accurately and with higher resolution.

\subsection{Overview of the Present Study}

PKS0349-27 is one of the few objects which is close enough for such
detailed study and which also displays an extensive, highly structured
ionized gas region in its host galaxy.  In this paper we report new
imaging and spectroscopic data for this object which identify regions
of extreme kinematic disruption in the EELG of this galaxy's host. The
analysis of these data is described in sections~\ref{images} and
\ref{spectra}.  The implications of the observed kinematics and
diagnostic emission line ratios for the origin and
ionization/excitation of this galaxy's EELG are discussed in
sections~\ref{velkin}, \ref{elr}, and \ref{jets}.

\section{Observations and Data Analysis} 

As part of a program to examine the complex kinematics in FRII radio
galaxies we obtained narrow band H$\alpha$ and continuum images and
long slit spectra of PKS0349-27 with the CTIO 1.5m and 4m telescopes.
The record of these observations is summarized in Table~\ref{tbl-1}.

\placetable{tbl-1}
 
\subsection{Optical Images} \label{images}

Direct images of PKS0349-27 were taken in January 1988 with the CTIO
1.5m, 888$\times$800 TI CCD (binned to 444$\times$400) and $f/13.5$
cassegrain secondary (lines 1 to 4 in Table~\ref{tbl-1}).  The CCD
scale was 0.31\,${''{px.}^{-1}}$.  The galaxy was observed through
matched on-band off-band filters with the aim of minimizing PSF
differences between the image sets. The paired images were reduced
using the routines in the IRAF ccdred package  and STSDAS wfpc, crrej,
which did an excellent job of removing ``cosmic rays.'' The resulting,
composite continuum image is shown as a logarithmic grayscale plot in
Figure\,1a. The continuum image was fit with the STSDAS
isophote routine to produce a model continuum galaxy, which was then
subtracted from the H$\alpha$ image, yielding the model-subtracted
image shown (as grayscale) in Figure\,1b. This technique leaves
intact the continuum sources, such as stars and other galaxies, which
are not directly associated with the extended continuum of the host
galaxy, but improves visibility of faint emission features by
minimizing subtraction-induced noise.

\begin{figure*}\label{fig1}\label{fig2}
\vspace*{-2.0in}
\begin{center}
\epsffile{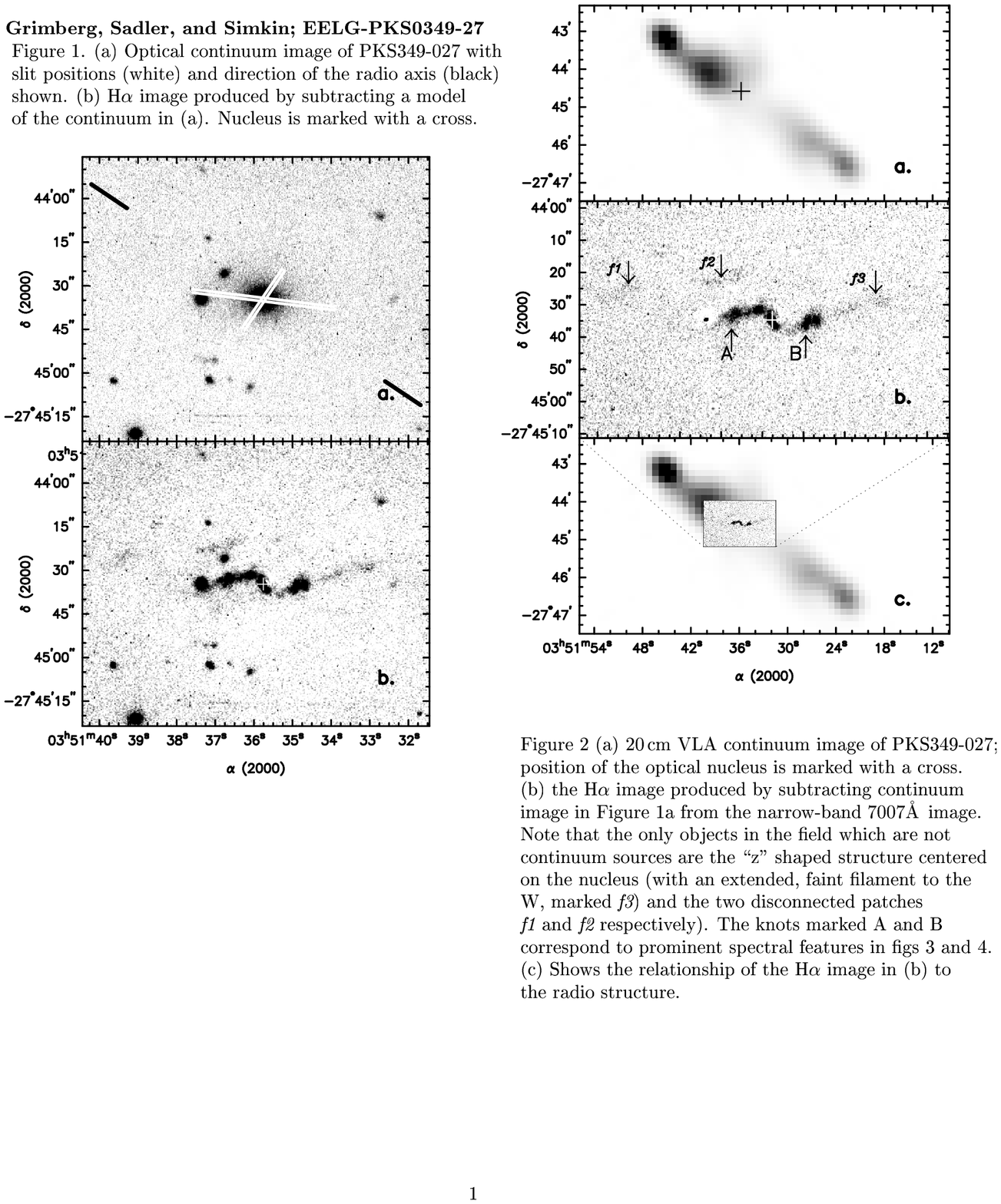}       
\end{center}
\end{figure*}

\placefigure{fig1}

A clearer picture of the relationship between the optical H$\alpha$
image and the radio structure is shown in Figure\,2.  Figure 2a
is a 20\,cm VLA radio continuum image (\cite{Si96}), and
Figure\,2b is the continuum subtracted (not model subtracted)
H$\alpha$ image. We note that all of the other ``sources'' in the field
subtract out in Figure\,2b, indicating that they are either
pure continuum sources  or emission line sources at a redshift which
differs from that of PKS0349-27 by at least 1700\,km\,s$^{-1}$.
Figure\,2c is a superposition of the H$\alpha$ image on the
20\,cm image with the galaxy nucleus  marked as a cross. The host
galaxy is clearly displaced towards the brighter (NE) radio lobe, which
has a hot spot that is seen in the 4.89 GHz map published by
\cite{Ek89}.  The faint arcs (marked as {\it f1, f2,} and {\it f3} in
Figure\,2b)  are also present in the [OIII] image of
\cite{Ha87}. The feature designated {\it f2} corresponds to the region
designated ``C'' by \cite{Ro87}. We note that the features {\it f1} and
{\it f2} appear to be parallel to the polarization vectors seen in the
20\,cm VLA map of \cite{Ba88}. Although this alignment may not be
significant, it does suggest the possibility that the extended radio
plasma may be interacting with the ionized gas in this region.

\placefigure{fig2}

\subsection{Optical Spectroscopy} \label{spectra}

Three long--slit spectra were taken during two consecutive nights in
February 1987 with the CTIO 4m spectrograph and the TI CCD .  All
exposures were bracketed with spectra of a HeArNe arc lamp taken with
the spectrograph in the same PA as the galaxy spectrum and the
telescope at the sky positions corresponding to the start and end
positions for the intervening galaxy spectrum.  The slit positions and
observing details  are shown in lines 5 through 7 in
Table~\ref{tbl-1}.  In all cases, the scale along the slit was 0.73\,$''{px}^{-1}$ and the slit width was 180 $\mu$m (1.2\,$''$). Slit
positions, widths and coverage are shown superimposed on the continuum
image in Figure\,1a.  Calibration stars (LTT1788 and LTT3864)
were taken at the beginning and end of each night.

The slit in PA 83$^{o}$ covers the extended arm-like features seen in
the host (shown in Figure\,1b) but does not lie along the radio
axis, which is, instead, aligned along the inner bar-like structure near
the nucleus.  The direction of the radio axis is indicated by the dark
bar in ~PA 56$^{o}$ in Figure\,1a.  At radial distances which
lie between 7 and 14 kpc\,${h^{-1}}$ from the nucleus, the spectra for
PA 83$^{o}$ exhibit strong bifurcation in the spatially extended
emission lines.  An image of the stronger spectral features for the
extended emission line regions in this PA  is shown in
Figure\,3.  The extreme distortions show up in both the
permitted lines  of H$\alpha$ (Figure\,3b) and the forbidden
lines of [OIII] and [NII] (Figure\,3a and b).  Knots labeled A
and B in Figure\,2b correspond to the  bifurcated spectral
lines in Figure\,3, which are marked A1 and B2. These are found
at velocities which are close to the nuclear velocity and hereafter
will be designated ``lo-vel'' clouds. The regions in the spectral lines
marked A2 and B1  are found at velocities with much higher amplitudes
relative to the nucleus, and hereafter will be designated ``hi-vel''
clouds.  Since these velocity anomalies appear in both the ``red''
spectra as well as the ``green'' spectra which were taken on different
nights, they are certainly real.

\begin{figure*}\label{fig3}\label{fig4}
\vspace*{-2.0in}
\begin{center}
\epsffile{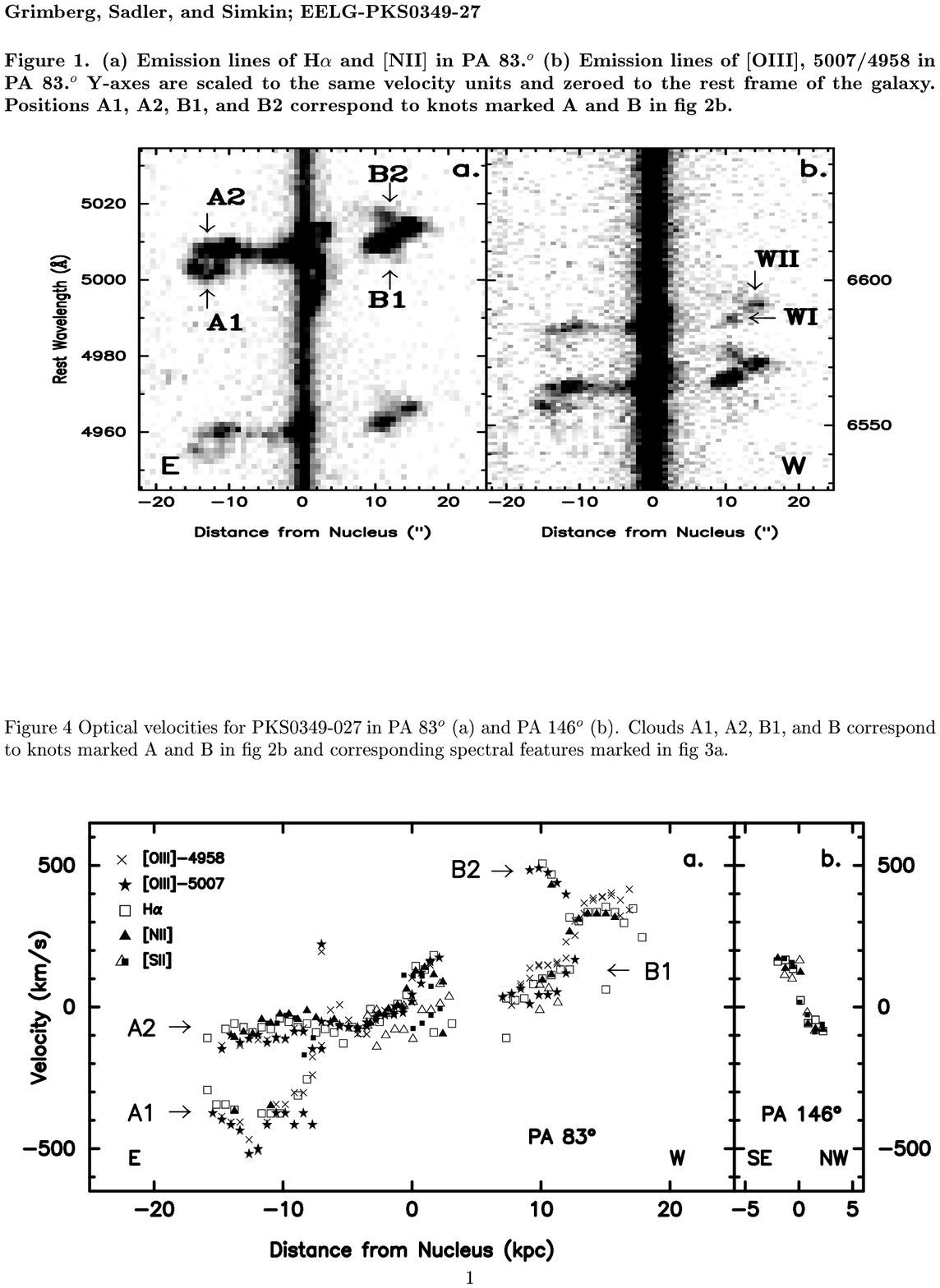}       
\end{center}
\end{figure*}

\placefigure{fig3}

\subsubsection{Velocities} \label{velval}

The spectra were reduced using a set of interactive routines (written
by SMS). They were flat fielded, calibrated in wavelength, calibrated
for S-distortion, which was minimal, and rebinned to remove
S-distortion and to produce constant wavelength increments.  A sinc
function was used for this rebinning. This is the proper
interpolation function for this type of evenly sampled spectral data,
since it mimics the slit function and has a modulation transfer
function tailored to finite data arrays.  The wavelength ranges and
dispersions are given in lines 5-7 of Table~\ref{tbl-1}.

The internal consistency in the  fits to the calibration spectra was
roughly 0.25 pixels, corresponding to $\pm 0.37\,$\AA\, for the
$\lambda\lambda$ 4895-5745 spectrum and $\pm 0.30\,$\AA\, for the
$\lambda\lambda$ 6653-7338 spectra. These translate into velocity
errors of $\sim$21\,km\,s$^{-1}$ and $\sim$14\,km\,s$^{-1}$
respectively.  Emission line positions in the calibrated spectra were
measured using an interactive routine which was adapted from analysis
routines developed by K. Ford at Carnegie Institution of Washington.
This routine finds the line peak, calculates the line centroid and the
center of a gaussian fit to its profile after background subtraction,
and automatically steps onto the next radial distance along the
direction perpendicular to the slit until the selection criteria are no
longer met.  The routine allows interactive setting of selection
criteria and baseline intervals. The velocities reported here are based
on measurements which have line strengths in excess of 10 times the rms
in the baseline background and which agree to within 0.5 pixels between
emission line positions determined by the two positioning routines.
While analyzing the data, it became clear that the emission lines
closer to the nucleus were composed of multiple peaks, representing at
least 2  separate velocity systems.  These peaks were measured using
line plots and an interactive cursor and are discussed in the next
paragraph. The error for these latter measurements cannot be determined
but is clearly greater than that for wavelength calibration fits and
probably greater than that for line fitting further out from the
nucleus.

The systemic velocity from these measurements, obtained from the
strongest emission lines -- [SII], [NII], H$\alpha$, and [OIII] -- is
19846$\pm$72\,km\,s.$^{-1}$ This is in surprisingly good agreement with
the value of 19846$\pm$130\,km\,s$^{-1}$ found by \cite{SB96} with
lower resolution spectra.  The error of $\pm$72\,km\,s,$^{-1}$ however,
far exceeds the value of $\sim$ 14 to 21 km$s^{-1}$ for the calibration
spectra and that of $\pm 21$ km$s^{-1}$ observed for other emission
line galaxies measured with the same equipment over two different
seasons (cf. \cite{Si99}).  Inspection of the nuclear spectra shows
that there are, in fact, at least two separate velocity systems; at
19840 $\pm 14$ km$s^{-1}$ and 19908 $\pm 18$ km$s^{-1}$) and possibly a
third, at 19708  km$s^{-1}$ seen only in the [SII] lines.  Thus, the
velocity measurements reported here have been zeroed to a nuclear
redshift velocity of 19800\,km\,s$^{-1}$  for purposes of determining
the internal kinematics.

The measured velocities are plotted in Figure\,4. The error
bars have been omitted in this plot to allow cleaner delineation of the
different line species. The scatter in the nuclear velocity
measurements noted above is clearly seen in Figure\,4a and
appears to be associated with the velocity bifurcations seen at larger
radii. The separate velocity systems are also seen in the PA 146$^{o}$
spectrum  but are easily lost in the steep velocity gradient across the
nucleus in that position angle (Figure\,4b). This suggests that
the dispersion in cloud velocities seen in the nuclear spectrum in both
position angles is more likely the result of limited spatial resolution
than true radial superposition of differentially moving clouds. The
bifurcated velocities for the clouds labeled A1, A2, B1, and B2 in
Figure\,3a are marked in Figure\,4a.

\placefigure{fig4}

\subsubsection{Emission Line Ratios} \label{ratios}

PA 83$^{o}$ was the only position angle  which was observed in {\it
both} the red and the blue-green spectral region. The line ratios from
these spectra for the emission lines [OIII] \,5007/H$\beta$, [SII]\,
6716+6731/H$\alpha$, and [NII]\, 6583/H$\alpha$ are shown in
Table~\ref{tbl-2}. In most cases only lines with signal to noise
(S/N)\, $\ge$\,3 were measured. In a few cases this criterion was
relaxed to S/N\,$\simeq$\,2. All errors listed in Table~\ref{tbl-2}
reflect 1$\sigma$ fluctuations from the measured line strengths. We
note, however, that an occasional single  emission line measurement can
have an erroneous value which is well outside this error limit if it
has been contaminated by a weak ``cosmic ray'' hit.  The [OI] 6300
emission line is seen not only in the nuclear spectrum of this object,
but is also visible as an extended feature, with the same
characteristic shape as the stronger emission lines shown in
Figure\,3. Although detectable as a continuous feature in video
images of the spectrum (and recorded by \cite{Ro87}), this line is far
too weak in our observed ccd spectrum to meet the  criteria imposed on
the  measurements reported in Table~\ref{tbl-2}.

\placetable{tbl-2}

Although we observed flux calibration standards for all of these
spectra, it is clear from examination and comparison of these
calibration spectra from night to night that the nights were
non-photometric, with flux variations of a factor of 2 or 3 from one
night to the next and up to 1.5 on the same night. Thus it is not
possible to obtain meaningful absolute values for the emission line
strengths nor is it possible to obtain meaningful flux ratios for
emission lines, such as [OI]$\lambda$\,6300 and [OIII]]$\lambda$\,5007,
which were taken with different grating settings on different nights.
We have, however, determined the {\it relative} corrections to the
ratios in Table~\ref{tbl-2} using the published absolute flux values
for the stars observed.  Even in the most extreme case of
[O\,I]/H$\alpha$ these are less than $\pm$\,0.04 in the log.  Since
this is much less than the errors from ``noise'' for most of the
observed emission lines, any error in the {\it relative} calibration
will be dominated by the emission line measurement errors themselves.
We reiterate here the important observational point made by
\cite{VO87}: Absolute flux measurements are very difficult to
accomplish, while diagnostic line ratios for adjacent emission lines
observed simultaneously can be very accurate.

The measured line ratios in Table~\ref{tbl-2} are plotted in a series
of ``BPT'' diagrams (cf. \cite{BPT81}, and \cite{VO87}), shown in
Figures\,5, \ref{fig7}, and \ref{fig8}. These plots will be
described and discussed in detail in section~\ref{elr}.  In addition to
the logarithmic line ratios in Table~\ref{tbl-2}, we have calculated
the line ratios for the [SII] $\lambda\lambda$ 6716/6731 doublet which
can be used to estimate the gas cloud densities. Although the errors
are substantial for any one value of this ratio, the general {\it
trends} in this density indicator are significant. If the EELG is
optically thin and has a constant temperature, a trend toward higher
values of the line ratio implies a trend towards lower values in the
gas density (cf. \cite{os89}). However, \cite{Fe97} have found that for
clouds with stratified densities and temperatures this line ratio may
not be useful as a density indicator.  The [SII] line ratios are
plotted, with low and high density limits marked, in Figure\,6a
and they are discussed in section~\ref{sii} below.

\section{Discussion} \label{discuss}

As noted in section~\ref{EELG} there are several different scenarios
involving the formation and fueling of radio galaxies which may be
relevant to an interpretation of the EELG in this object. In broad
outline they are:

\begin{enumerate}

\item Possible galaxy collisions, mergers, and tidal forcing of the
extended gas by a nearby companion.

\item Ionization and excitation of the EELG by continuum radiation from
the nuclear ``engine''.

\item  Interaction between a radio plasma jet and the interstellar
medium in the host galaxy.

\end{enumerate}

All of these possibilities have been discussed for PKS0349-27 in
earlier papers (cf. \cite{Da84,Ro87,Ta89,Vi92,SB96} and \cite{Wi97}).
However, the present data, with their better spectral resolution,
provide additional information with new implications for these various
scenarios. 

\subsection{Morphological Structure of the Ionized Gas - Tides?} \label{morph}
There are several studies in the literature which support the  idea
that AGNs are fueled by infalling material which has been channeled
into a galaxy's nuclear regions by a merger or other tidal disturbance
(cf. \cite{mac94} and references therein).  However, the short time scales
involved in this process and the fact that some  mergers and tidal
disturbances {\it do not} fuel AGNs make it difficult to assess the
importance of this mechanism for radio galaxies. There are several known
examples of candidates for galaxy mergers in non-radio galaxies which
show that the extended ionized gas in these systems is photoionized by
starburst activity (\cite{cr89}, \cite{cr94} , \cite{hi94},
\cite{hi96}, \cite{ar97}). Thus, starbursts may be one way of
identifying merger activity in AGN hosts.

The long, two-armed structure of the ionized gas, shown in
Figures\,1b and \ref{fig2}b, and the [OIII] images of
\cite{Ha87} and \cite{Ba88}, is similar to the type of structure
associated with tidally interacting galaxies (\cite{To77,By90} and
references therein).  The suggestion of tidal interaction is reinforced
by the presence of a companion galaxy with a radial velocity within
$\sim$200\,km\,s$^{-1}$ of the PKS0349-27 host which is seen at the end
of the object's eastern arm in Figure\,1b. This was first
identified  by \cite{Da84}, who measured its redshift.  If the
structure of the EELG is determined by a tidal encounter or a merger,
then the most likely geometry is one in which we are viewing the orbit
of the merging companion nearly edge-on (cf. \cite{To77,By90}). Such
systems often have gas velocities which are bifurcated or counter
rotational (op cit.).

\subsection{Ionized Gas Velocities} \label{velkin}

The data plotted in Figure\,4a show that the most notable split
in velocities occurs for gas at R\,$>$\,7\,kpc\,${h^{-1}}$.  This type
of velocity detail was not identified in earlier measurements of this
object (\cite{Da84,Ta89,SB96}), either because it is peculiar to this
particular position angle, unresolved in the earlier spectra, or
because the line intensities in the low velocity system are weaker than
in the high velocity one.  Since the extreme velocity separation of
$\sim$1000\,km\,s$^{-1}$ between the systems in A1 and B2 is far in
excess of any velocity amplitudes reported by either \cite{Da84} or
\cite{Ta89} for the gas at  PA {90/270,}$^{o}$ it appears likely that
these  earlier data  refer to gas clouds associated with those labeled
A2 and B1 in Figures\,3 and \ref{fig4}.  Because these two
systems (A2 and B1) are also stronger (Figure\,3), the presence
of the weaker systems (A1 and B2) in other position angles cannot be
ruled out by these earlier measurements. Finally, although the split in
velocity systems is most notable for the knots labeled A and B in
Figure\,2b, the velocity points in Figure\,4a as well
as the discussion in section~\ref{velval} show that this separation in
cloud velocities extends into regions closer to the nucleus, but with
much smaller amplitude. Thus, there is a strong possibility that {\it
both} velocity systems are present throughout the host galaxy but the
weaker one has been undetected until now.

\cite{Da84} present a composite rotation-expansion model for the gas in
the host of PKS349-27. Although the ``rotation'' velocities in
Figure\,4b (PA 146$^{o}$) are roughly consistent with that
earlier model, the bifurcated velocities shown in Figure\,4a
(PA 83$^{o}$) show that the situation is much more complex than simple
rotation-expansion.  As a first step in resolving this complexity, we
believe it is important to try to determine which of the different
velocity clouds are physically associated.  The two most likely
patterns for the gas velocities shown in Figure\,4a are: (1)
The clouds comprising sets A2 and B2 may be physically associated with
a high redshift (``hi-z'')  system with respect to the host nucleus;
while those in  A1 and B1 comprise a low redshift (``lo-z'') system.
(2) The extreme velocity clouds for the set A1 and B2 may be physically
associated with a system which has a high rotational velocity relative
to the host galaxy nucleus (``hi-vel'' clouds), while the clouds  (A2
and B1), comprise a separate, independent system with relatively low
rotational velocities (``lo-vel'' clouds).

\subsection{Line Ratios - Ionization and Excitation} \label{elr}
\begin{figure*}\label{fig5}\label{fig6}\label{fig7}\label{fig8}
\vspace*{-2.0in}
\begin{center}
\epsffile{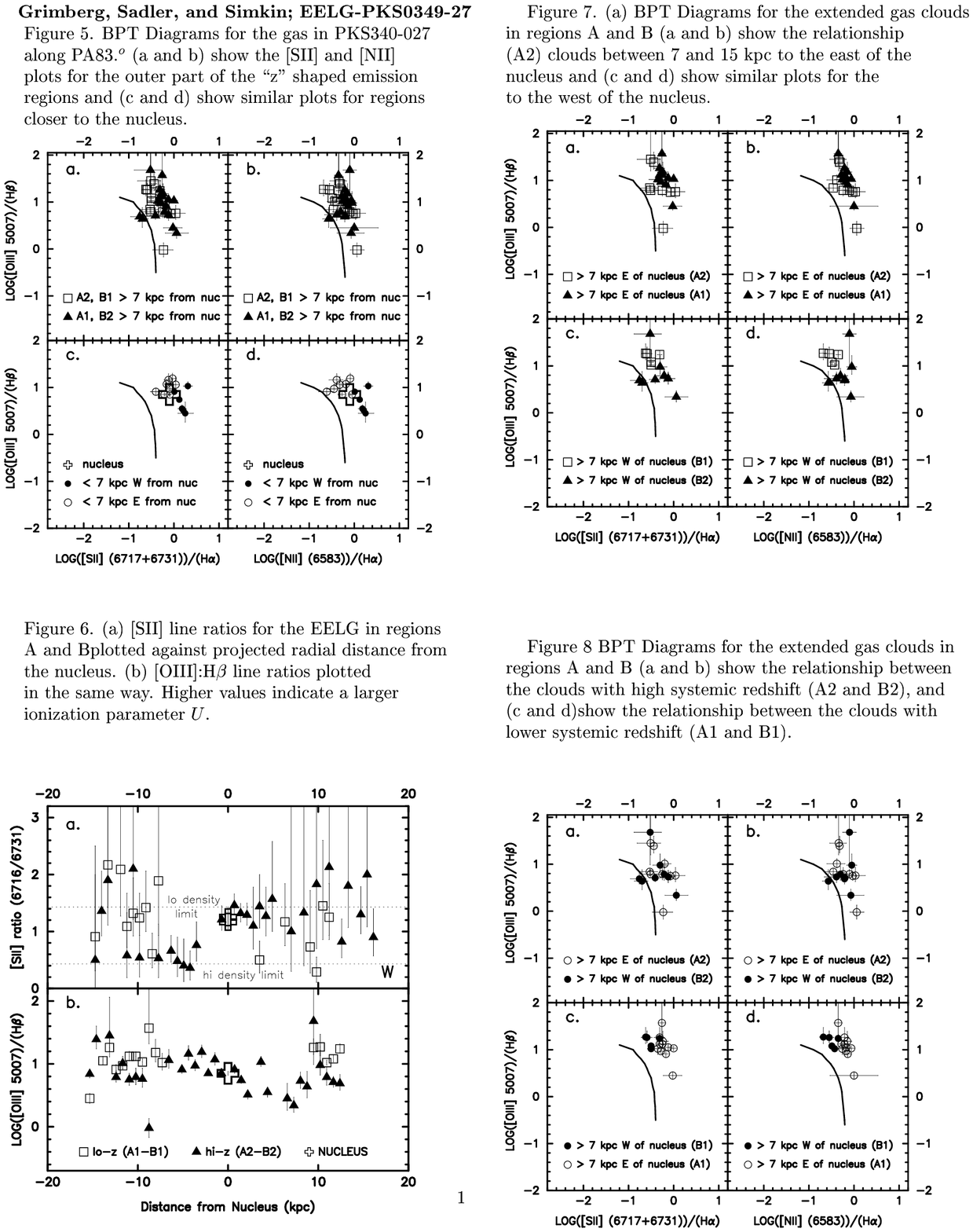}       
\end{center}
\end{figure*}

We have used the BPT plots of measured emission line ratios in
Figures\,5, \ref{fig7}, and \ref{fig8} to try to delineate the
relationship between the different cloud systems. The solid curves in
each diagram in Figures\,5 through \ref{fig8} are taken from
\cite{VO87},  and mark the empirical division between HII-region
like objects, which lie to the left hand side of the line, and objects,
such as AGNs, ionized by a harder radiation spectrum, which lie to the
right hand side of the line.

\subsubsection{Star Formation?} \label{star}

First, in an attempt to identify any gross differences in the physical
conditions between the EELG, we have separated the emission line
measurements into two different sets according to radial distance from
the nucleus. Those with projected distances $>$\,7 kpc\,${h^{-1}}$ from
the  nucleus (Figures\,5a and b) are plotted with the
``hi-vel'' clouds (A1 and B2) as filled symbols and the ``lo-vel''
clouds as open symbols.  Those with projected distances $<$\,7
kpc\,${h^{-1}}$ from the  nucleus (Figures\,5c and d), are
plotted with points to the west of the nucleus marked as filled symbols
and  points to the east of the nucleus marked as open symbols.

Figure\,5 allows us to immediately test the hypothesis that one
of the velocity systems arises from gas which has been disturbed
(either by the radio plasma or as infall from a merging galaxy) and the
other originates in a more ``normal,'' gaseous component, ionized by
recent starbursts.  The results of this test are clear: In
Figure\,5 {\it both} systems exhibit the high excitation
conditions which are the signature of ionization by a strong UV
continuum or by shock heating.  Although there is more scatter in the
points from regions $>$\,7 kpc\,${h^{-1}}$ from the nucleus, all fill a
similar region of the BPT diagrams, and all lie to the left (the AGN
region) of the dividing line between AGN spectra and HII regions.  Thus
our first conclusion from this data is that there is no detectable
starburst activity in the EELG.

\placefigure{fig5}

\subsubsection{Ionization/Excitation from a Central Source?}  \label{sii} 

The diagrams in Figure\,5 show no significant change in
the line ratios as a function of distance.
This is in general agreement with the conclusions of \cite{Da84} and
\cite{Ro87} that there is no appreciable change in the ionization
parameter, $U$, of the EELG with radial distance from the central
nuclear source. The conclusion they drew from this fact was that the
gas density must decrease with radial distance. (Since $U =
\frac{Q}{4\pi\,c\,R^{2}n_H}$, where $Q$ is the nuclear ionizing photon
luminosity, $R$ the radial distance from the nucleus, and $n_H$, the
hydrogen number density, then if $U$ is constant $n_H$ must decrease.)
However, a closer look at the BPT diagrams in Figure\,5 and the
[SII] plots in  Figure\,6a for points  $<$\,7\,kpc\,${h}^{-1}$
shows that the situation is more complex than this. The trend toward
{\it increasing} density toward the east (seen in Figure\,6a)
does not fit into this picture. An increased density should lead to a
{\it decreasing} ionization parameter (decreasing [OIII]/H$\beta$) in
this direction, but the opposite is seen in Figure\,5c and d.
On the other hand, the clouds on the {\it western} side of the nucleus
do seem to fit the central source picture. They tend to have lower
ionization (lower [OIII]/H$\beta$) than those on the eastern side, and
the increase in the [SII] ratios indicates that the clouds to the west
also have decreasing density, balancing out the increase in $R$.

To help clarify this, the parameter which is most closely correlated
with ionization ([OIII]/H$\beta$, cf. \cite{VO87}), is plotted as a
function of radial distance in Figure\,6b.  Comparison of
Figures\,6a and b in the range -7\,$\leq\,R\leq$\,7 shows that
the trends predicted by a simple nuclear source photoionization model
do not fit the observed density trends.  These trends are significant
within the measured errors and also consistent with the H$\alpha$ image
of the gas (Figures\,1b and \ref{fig2}b), which is fainter in
the region to the west of the nucleus, showing that the gas there has a
lower emission measure than the gas on the eastern side. We must
conclude that either the gas on the western side of the EELG has
$n_H\,\sim\,R^{-2}$ with the number of ionizing photons falling off as
$R^{-2}$ while the eastern side is somehow receiving more ionizing
photons at greater distances from the nucleus, or the eastern EELG
clouds have an additional source of ionization/excitation which
supplements  the nuclear source.

However, one interesting insight into the physical state of the EELG at
$R\,\ge$\,7\,kpc$\,{h}^{-1}$ can be obtained from the simple central source
model. Figure\,6 shows that the lo-z system has lower densities
and a higher $U$ (higher [OIII]:H$\beta$) than the hi-z system.  Thus,
{\it if} the EELG is ionized by a central source,  the relationship
between $U$ and $n_{H}$ (given above) suggests that the two sets of
clouds may well be at the same physical radius as well as the same
projected radius from the nucleus.

While the above discussion is consistent with conclusions of
\cite{Ro87},  \cite{SB96}, and \cite{Wi97}, where it was noted that
different measured line ratios in PKS0349-27  were inconsistent with a
global picture of ionization/excitation in the EELG based on
photoionization by the (hidden) nuclear continuum, our findings are
much more detailed and direct.  The  radial distributions of
density and ionization reported here,  coupled with the different
characteristics of the two velocity systems seen in PA 83,$^{o}$
strongly suggest that whatever the final interpretation of this object,
it must take the velocity systems into account.

In summary, these discrepancies in applying a central source model to
the observed conditions in this galaxy's EELG when combined with the
observation by \cite{Zi98} that its nucleus has little or no observed
UV flux, implies that the  source of at least some of the
ionization/excitation observed in the extended gas has a non-nuclear
origin.

\placefigure{fig6}

\subsubsection{Relationship Between the Different Velocity Systems?}
\label{sys}

In an attempt to identify common factors amongst gas clouds with
$R\,\ge$\,7\,kpc\,${h}^{-1}$ we have sorted them into the different
velocity sets defined in section~\ref{velkin} (hi-lo-vel and hi-lo-z)
and plotted the BPT diagrams shown in Figures\,7 and
\ref{fig8}. The plots with the most overlap (Figure\,8)  seem
to be the ones in which the line ratios for the lo-z (A1 and B1) clouds
are plotted together (Figure\,8c and d). Almost all of these
have values of [OIII]/H$\beta$\,$\geq$10. The hi-z clouds (A2 and B2,
Figure\,8a and b) also show good overlap and a slightly lower
[OIII]/H$\beta$ value ($\sim$\,8) but with more scatter. What this
means physically is that the gas in the lo-z clouds has a slightly
higher ionization/excitation state than that in the hi-z clouds. This
is consistent with the [SII] line ratio plots in Figure\,6a
which indicate that the lo-z clouds have consistently lower densities,
while the hi-z clouds show more scatter (see section~\ref{sii} above).

If these similarities in ionization and density imply physical
association, then the velocities in Figure\,4a can be
reconciled in the following way: Rotating the curves by 180$^{o}$ and
shifting them in the radial direction (X-axis) by $\sim$\,2.9 kpc shows
that the hook in the velocity curve at A1 and the flat region at B1
appear to be mirror images of the upper branch in the velocities at B2
and the flat part of the velocity plot at A2. However, this
transformation  does not fit the velocity points traced by the [SII]
measurements, which we noted in Section~\ref{velval} appeared to have three
velocity components near the nucleus. These [SII] velocities seem
to trace a  system with a slightly negative velocity slope near the
nucleus. Both of these features, asymmetrically shifted velocity curves
which are mirror images, and small, inner systems which show counter
rotation, are found in merging systems and tend to support the
``merger'' scenario for the origins of the gas (cf. \cite{mac94} and
references therein).

\placefigure{fig7}
\placefigure{fig8}

\subsection{Radio Jets and the EELG?} \label{jets}

Although the symmetric velocity structure described in
section~\ref{sys} above may be consistent with  a
tidal-interaction/merger, the magnitude of the velocity separations
between the two cloud systems is also consistent with the pattern of
disruption caused by the passage of a radio jet \footnote{We note that
the ``jet'' seen at $\sim$PA 56$^{o}$ in the original identification
images by \cite{Se68} is most likely the inner part of the extended
ionized gas seen in Figure\,1b.}.  The time scale for  a
disturbance in the gas to propagate from the location of the radio axis
(at PA\,56$^{o}$) to the observed position in the ionized gas at
PA\,83$^{o}$ when the gas is moving at a relative velocity of
$\sim$350\,km\,s,$^{-1}$ is $\sim$10$^{7}$ years. This is comparable to
the  age of this type of FRII source (6-30$\times\,10{^{6}}$an, \cite{Ca91}) and since the observed radial velocity differences
between regions B1 and B2 and A1 and A2 exceed $\sim$400\,km\,s,$^{-1}$
at points, we cannot completely rule out collisional ionization of the
EELG by the material ejected from the nucleus in a radio jet.  However,
there is some evidence that this type of jet-cloud collision generates
ionization/excitation conditions which differ from both those for
AGN-like gas and HII regions (\cite{SGS99}).  In the few clear examples
of an immediate jet-cloud collision (NGC\,7385, \cite{Si84}, and Pictor
A, \cite{Si99}), the signature of such jet-ionized gas appears to be
relatively low excitation (with a value of [OIII]/H$\beta\,\le$\,4)
along with AGN-like values of [SII]/H$\alpha$ and [OI]/H$\alpha$ but
relatively lower values of [NII]/H$\alpha$ (similar to those found for
HII regions). The observed line ratios in Table~\ref{tbl-2} lie outside
these ranges. This does not completely rule out the possibility that a
past disturbance by a radio jet has ionized the gas in PKS0349-27, but
the recombination times for gas at the densities indicated by the [SII]
line ratios in Figure\,6a will be shorter than the propagation
time scale noted above and thus an additional source for the present
ionization is still necessary.

\section{Summary and Conclusions}

Although the material reported here provides a much more detailed
picture of the ionized gas in PKS0349-27 than previous observations, it
raises several new questions.

On the positive side, these measurements point up the importance of the
greater accuracy possible when diagnostic line ratios are analyzed in
terms of the BPT scheme proposed by \cite{VO87}. However, in spite of
the observational advantages of this system, the discussions in
sections~\ref{sii} and \ref{elr} make it clear that it is difficult to
compare the line ratio diagrams discussed here with either models or
data reported in the more common systems which rely on  [OI]:][OIII] or
[OII]:[OIII] criteria for ionization/excitation determinations.
Hopefully, there will be soon more detailed theoretical models which include published line ratios  calculated for the \cite{VO87} system.

The present measurements lead to the following conclusions:

\begin{enumerate}
\item The ionization of the EELG in PKS0349-27 does not come from starburst
activity.
\item The bifurcated velocity systems in PA\,83$^{o}$ of this system
appear to have similar emission line ratios (and thus similar
ionization parameters) when grouped into hi-z and lo-z systems.
\item There may be yet a third  velocity system near the nucleus of
this system which is in counter rotation.
\item The gas clouds at projected distances closer than
7\,kpc\,$h^{-1}$ to the nucleus of PKS0349-27 have line ratios which
are inconsistent with a simple central source ionization model and
(particularly on the eastern side) require either an additional
ionization source or a completely different mechanism which is
unrelated to the nuclear source.
\item The most likely conclusion from these facts is that the gas has
been disturbed by a merger or tidal interaction with a nearby galaxy,
but the possibility of disruption by a radio jet cannot be completely
ruled out.
\end{enumerate}

On the other hand, there are several immediate questions raised by these data. Among the most obvious are:
 
\begin{enumerate}
\item What is the detailed velocity field of the EELG in PKS0349-27? Is the weaker, high velocity system only seen in PA\,83,$^{o}$ or does it extend into all areas around the nucleus?

\item What is the source of the ``additional'' ionization/excitation on the E side of the host? How can we distinguish between shock heating due to the merger of two galaxies and that caused by the passage of a radio jet?

\item What is the detailed radio structure (at arc second resolution) op
f PKS0349-27?  Are the ionized gas filaments seen to the east of the nucleus interacting with the radio plasma? If so, what are the implications for the ionization/excitation of the EELG?
\end{enumerate}

These questions underscore the need for new high resolution studies,
with improved equipment (both radio and optical) of not only this
object but also other nearby, FRII, radio galaxies for comparison.

\acknowledgments

This research has made use of the NASA/IPAC Extragalactic Database (NED)
which is operated by the Jet Propulsion Laboratory, California
Institute of Technology, under contract with the National Aeronautics
and Space Administration. The CTIO observations were done with some
support from NSF AST-89-14567 and publication costs from a grant
(to SMS) from NASA, administered by the American Astronomical
Society.  The graphics in this paper were generated with the software
package $WIP$\footnote{$WIP$ is copyright by the
Berkeley-Illinois-Maryland Association (BIMA) Project (\cite{Mo95})},
using the $PGPLOT$\footnote{software copyrighted by California
Institute of Technology} graphics library.

\begin{deluxetable}{lcccccr}
\footnotesize
\tablecaption{CTIO - Optical Images and Spectra \label{tbl-1}}
\tablewidth{0pt}
\tablehead{
\colhead{TELESCOPE} &\colhead{DATE }  &\colhead{EXP TIME} &\colhead{AIRMASS}  &\colhead{${\lambda}_{e}$(range)} &\colhead{$\Delta\lambda$} &\colhead{SEEING}} 

\startdata

CTIO-1.5m&22/23 Jan 1988 &4000 s&1.02 &7007\AA &78-line &1.34$''$\nl
CTIO-1.5m&22/23 Jan 1988 &4000 s&1.12 &6560\AA &110-cont &1.24$''$\nl
CTIO-1.5m&23/24 Jan 1988 &4000 s&1.03 &7007\AA &78-line &1.39$''$\nl
CTIO-1.5m&23/24 Jan 1988 &4000 s&1.42 &6560\AA &110-cont &1.78$''$\nl
CTIO-4m-PA 146$^{o}$&15/16 Jan 1988  &2000 s&1.30 &6653-7338 \AA&1.20 \AA/px &
1.43$''$\nl                     
CTIO-4m-PA 83.1$^{o}$&15/16 Jan 1988  &2000 s&1.56 &6653-7338 \AA&1.20 \AA/px &1.52$''$\nl                     
CTIO-4m-PA 83.1$^{o}$ &16/17 Jan 1988  &2000 s&1.19 &4895-5745 \AA&1.49 \AA/px &1.67$''$\nl                   

\enddata

\end{deluxetable}

\begin{deluxetable}{ccccccc}
\footnotesize
\tablecaption{ Emission Line Ratios in PA 83.1$^{o}$  \label{tbl-2}}
\tablewidth{0pt}
\tablehead{
\colhead{R (\,kpc\,${''h}^{-1}$)}
&\colhead{log($\frac{[N\,II]}{H\alpha}$)}
&\colhead{log($\frac{[S\,II]}{H\alpha}$)}
&\colhead{log($\frac{[O\,III]}{H\beta}$)}
&\colhead{log($\frac{[N\,II]}{H\alpha}$)}
&\colhead{log($\frac{[S\,II]}{H\alpha}$)}
&\colhead{log($\frac{[O\,III]}{H\beta}$)}}
\startdata

& A1 and B2& &&A2 and B1&&\nl
  -14.72&   0$^{+ 0.53}_{  -0.53}$&-0.02$^{+0.2}_{  -0.21}$&0.45$^{+ 0.08}_{  -0.08}$&
 -0.46$^{+ 0.15}_{  -0.19}$&-0.53$^{+ 0.17}_{  -0.22}$&0.84$^{+ 0.06}_{  -0.06}$\nl
  -14.02&\nodata&\nodata&\nodata&-0.32$^{+ 0.05}_{  -0.05}$&-0.44$^{+ 0.06}_{  -0.06}$&1.39$^{+ 0.21}_{  -0.15}$\nl
 -13.32&-0.17$^{+ 0.08}_{  -0.08}$&-0.12$^{+ 0.07}_{  -0.08}$&1.05$^{+ 0.07}_{  -0.06}$&\nodata&\nodata &\nodata \nl
  -12.61&-0.21$^{+0.1}_{   -0.1}$&-0.31$^{+ 0.11}_{  -0.12}$&1.26$^{+ 0.21}_{  -0.15}$&
 -0.34$^{+ 0.17}_{   -0.2}$&-0.51$^{+ 0.22}_{   -0.3}$&1.45$^{+ 0.61}_{  -0.25}$\nl
  -11.91&-0.14$^{+ 0.08}_{  -0.08}$&-0.17$^{+ 0.08}_{  -0.09}$&0.91$^{+ 0.08}_{  -0.07}$&
-0.1$^{+0.1}_{  -0.11}$&-0.51$^{+ 0.18}_{  -0.23}$&0.79$^{+0.1}_{  -0.08}$\nl
 -11.21&-0.18$^{+ 0.06}_{  -0.06}$&-0.28$^{+ 0.07}_{  -0.07}$&0.97$^{+ 0.06}_{  -0.05}$
&-0.38$^{+ 0.18}_{  -0.22}$&-0.2$^{+ 0.14}_{  -0.15}$&1.01$^{+ 0.11}_{  -0.09}$\nl
  -10.51&-0.28$^{+ 0.06}_{  -0.06}$&-0.31$^{+ 0.06}_{  -0.06}$&1.12$^{+ 0.08}_{  -0.07}$
&-0.03$^{+0.1}_{   -0.1}$&-0.07$^{+0.1}_{  -0.11}$&0.75$^{+0.1}_{  -0.09}$\nl
  -9.81&-0.21$^{+ 0.04}_{  -0.05}$&-0.25$^{+ 0.05}_{  -0.05}$&1.12$^{+ 0.07}_{  -0.06}$
&-0.24$^{+ 0.12}_{  -0.13}$&-0.25$^{+ 0.12}_{  -0.13}$&0.79$^{+ 0.14}_{  -0.12}$\nl
   -9.11&\nodata&\nodata&\nodata&0.03$^{+ 0.22}_{  -0.21}$&0.04$^{+ 0.21}_{  -0.21}$&0.76$^{+ 0.17}_{  -0.13}$\nl
   -8.41&0.06$^{+ 0.16}_{  -0.15}$&-0.23$^{+ 0.21}_{  -0.24}$&-0.02$^{+ 0.15}_{  -0.15}$
1&-0.35$^{+ 0.08}_{  -0.09}$&-0.26$^{+ 0.08}_{  -0.08}$&1.57$^{+ 0.61}_{  -0.25}$\nl
   -7.71&-0.15$^{+0.1}_{  -0.11}$&-0.24$^{+ 0.12}_{  -0.12}$&1.18$^{+ 0.21}_{  -0.15}$&\nodata&\nodata &\nodata \nl
   -7.01&-0.26$^{+ 0.15}_{  -0.17}$&-0.35$^{+ 0.17}_{   -0.2}$&1.02$^{+ 0.15}_{  -0.12}$&\nodata&\nodata &\nodata \nl
   -6.31&-0.32$^{+ 0.12}_{  -0.13}$&0.04$^{+ 0.08}_{  -0.08}$&1.06$^{+ 0.17}_{  -0.13}$&\nodata&\nodata &\nodata \nl
   -4.91&-0.61$^{+ 0.14}_{  -0.18}$&-0.4$^{+0.1}_{  -0.12}$&0.91$^{+ 0.07}_{  -0.06}$&\nodata&\nodata &\nodata \nl
    -4.2&-0.39$^{+ 0.14}_{  -0.17}$&-0.12$^{+0.1}_{   -0.1}$&1.16$^{+ 0.13}_{   -0.1}$&\nodata&\nodata &\nodata \nl
    -3.5&-0.45$^{+ 0.14}_{  -0.16}$&-0.09$^{+ 0.09}_{  -0.09}$&0.97$^{+0.1}_{  -0.08}$&\nodata&\nodata &\nodata \nl
    -2.8&-0.09$^{+ 0.08}_{  -0.08}$&-0.03$^{+ 0.07}_{  -0.07}$&1.19$^{+ 0.11}_{  -0.09}$&\nodata&\nodata &\nodata \nl
    -2.1&-0.27$^{+ 0.08}_{  -0.09}$&-0.21$^{+ 0.08}_{  -0.08}$&0.85$^{+ 0.04}_{  -0.03}$&\nodata&\nodata &\nodata \nl
    -1.4&-0.19$^{+ 0.05}_{  -0.05}$&-0.16$^{+ 0.05}_{  -0.05}$&1.07$^{+ 0.11}_{  -0.09}$&\nodata&\nodata &\nodata \nl
    -0.7&-0.05$^{+ 0.03}_{  -0.03}$&-0.09$^{+ 0.03}_{  -0.03}$&0.85$^{+ 0.04}_{  -0.04}$&\nodata&\nodata &\nodata \nl
     0  &-0.05$^{+0.05}_{  -0.05}$&-0.06$^{+0.04}_{  -0.04}$& 0.87$^{+ 0.07}_{  -0.03}$&\nodata&\nodata &\nodata \nl
     0.7&-0.03$^{+ 0.03}_{  -0.03}$&0.01$^{+ 0.03}_{  -0.03}$&0.91$^{+ 0.06}_{  -0.06}$&\nodata&\nodata &\nodata \nl
     1.4&-0.02$^{+ 0.03}_{  -0.03}$&0.12$^{+ 0.03}_{  -0.03}$&0.74$^{+ 0.06}_{  -0.05}$&\nodata&\nodata &\nodata \nl
     2.1&0.02$^{+ 0.05}_{  -0.05}$&0.21$^{+ 0.04}_{  -0.04}$&0.51$^{+ 0.07}_{  -0.07}$&\nodata&\nodata &\nodata \nl
    6.31&0.08$^{+ 0.21}_{   -0.2}$&0.25$^{+ 0.18}_{  -0.16}$&0.45$^{+ 0.24}_{  -0.19}$&\nodata&\nodata &\nodata \nl
     3.5&-0.27$^{+ 0.18}_{  -0.21}$&0.31$^{+0.1}_{   -0.1}$&1.03$^{+ 0.09}_{  -0.07}$&\nodata&\nodata &\nodata \nl
     4.2&-0.22$^{+ 0.13}_{  -0.15}$&0.18$^{+ 0.09}_{  -0.09}$&0.55$^{+ 0.08}_{  -0.08}$&\nodata&\nodata &\nodata \nl
    7.01&-0.07$^{+ 0.29}_{  -0.31}$&0.06$^{+ 0.26}_{  -0.25}$&0.34$^{+ 0.13}_{  -0.11}$&\nodata&\nodata &\nodata \nl
    7.71&-0.39$^{+ 0.21}_{  -0.27}$&-0.12$^{+ 0.15}_{  -0.16}$&0.73$^{+ 0.14}_{  -0.12}$&\nodata&\nodata &\nodata \nl
    8.41&-0.57$^{+ 0.11}_{  -0.13}$&-0.7$^{+ 0.13}_{  -0.17}$&0.64$^{+ 0.24}_{  -0.18}$&\nodata&\nodata &\nodata \nl
    9.11&-0.1$^{+ 0.15}_{  -0.15}$&-0.52$^{+ 0.25}_{  -0.36}$&1.68$^{+ 0.69}_{  -0.69}$
 &-0.55$^{+0.1}_{  -0.11}$&-0.59$^{+ 0.11}_{  -0.13}$&1.26$^{+ 0.14}_{  -0.11}$\nl
   9.81&-0.05$^{+ 0.11}_{  -0.11}$&-0.3$^{+ 0.15}_{  -0.16}$&0.98$^{+ 0.24}_{  -0.16}$
&-0.68$^{+ 0.07}_{  -0.08}$&-0.62$^{+ 0.07}_{  -0.07}$&1.27$^{+0.2}_{  -0.14}$\nl
   10.51&-0.3$^{+ 0.16}_{  -0.18}$&-0.21$^{+ 0.14}_{  -0.16}$&0.79$^{+ 0.16}_{  -0.13}$
&-0.43$^{+ 0.06}_{  -0.07}$&-0.5$^{+ 0.07}_{  -0.08}$&1.02$^{+ 0.04}_{  -0.04}$\nl
   11.21&-0.2$^{+ 0.09}_{  -0.09}$&-0.41$^{+ 0.11}_{  -0.13}$&0.71$^{+ 0.09}_{  -0.08}$
 &-0.49$^{+ 0.05}_{  -0.05}$&-0.5$^{+ 0.05}_{  -0.05}$&1.08$^{+ 0.06}_{  -0.05}$\nl
   11.91&-0.21$^{+ 0.06}_{  -0.07}$&-0.76$^{+ 0.15}_{   -0.2}$&0.69$^{+ 0.14}_{  -0.11}$
  &-0.35$^{+ 0.05}_{  -0.05}$&-0.31$^{+ 0.05}_{  -0.05}$&1.24$^{+ 0.08}_{  -0.07}$\nl
   12.61&\nodata&\nodata&\nodata&-0.29$^{+ 0.09}_{   -0.1}$& 0$^{+ 1.1}_{     - 0}$&0.99$^{+ 0.11}_{  -0.09}$\nl

\enddata
\end{deluxetable}

\clearpage
\oddsidemargin -1.0in
\evensidemargin -1.0in
\textwidth 7.5in

\clearpage

\clearpage

\clearpage
\end{document}